\documentclass[fleqn,10pt]{wlscirep}
\title{Carnot efficiency is reachable in an irreversible process}

\author[1,*]{Jae Sung Lee}
\author[1,+]{Hyunggyu Park}
\affil[1]{School of Physics and Quantum Universe Center, Korea Institute for Advanced Study, Seoul 02455, Korea}
\affil[*]{jslee@kias.re.kr}
\affil[+]{hgpark@kias.re.kr}

\begin{abstract}
In thermodynamics, there exists a conventional belief that ``the Carnot efficiency is reachable only when a process is reversible.'' However, there is no theorem proving that the Carnot efficiency is unattainable in an irreversible process. Here, we show that the Carnot efficiency is reachable in an irreversible process through investigation of the Feynman-Smoluchowski ratchet (FSR). We also show that it is possible to enhance the efficiency by increasing the irreversibility.
Our result opens a new possibility of designing an efficient heat engine in a highly irreversible process and also answers the long-standing question of whether the FSR can operate with the
Carnot efficiency.
\end{abstract}
\begin{document}

\flushbottom
\maketitle
% * <john.hammersley@gmail.com> 2015-02-09T12:07:31.197Z:
%
%  Click the title above to edit the author information and abstract
%
\thispagestyle{empty}

%\noindent Please note: Abbreviations should be introduced at the first mention in the main text – no abbreviations lists. Suggested structure of main text (not enforced) is provided below.

\section*{Introduction}

Thermodynamics is a field of science dealing with the relationship between energy, work, and heat~\cite{Fermi}. It was practically initiated to develop a heat engine with high efficiency. Here, the heat engine is a device transforming heat energy into useful mechanical work. Therefore, one of the most interesting subjects in thermodynamics is the study of the maximum possible efficiency attainable by a heat engine. The maximum efficiency of a heat engine operating in two thermal baths of different temperatures $T_1$ and $T_2$ ($T_1>T_2$) is fairly well understood; the efficiency cannot be greater than the Carnot efficiency $\eta_\textrm{C}=1-T_2 / T_1$~\cite{Kittel}.

The efficiency can reach $\eta_\textrm{C}$ when the process of the heat engine is perfectly reversible~\cite{Kittel}. Formally, defining $\mathcal{Q}_1$ and $\mathcal{Q}_2$ as the average heat transferred from thermal baths at temperatures $T_1$ and $T_2$ during one engine cycle over a time duration $\tau_\textrm{cyc}$, respectively, then, the efficiency $\eta$ and the entropy production per cycle $\Delta S$ are defined as
\begin{eqnarray}
\eta &\equiv& 1 - \frac{| \mathcal{Q}_2 |}{|\mathcal{Q}_1|}, \nonumber\\
\Delta S &\equiv& -\frac{|\mathcal{Q}_1|}{T_1} + \frac{|\mathcal{Q}_2|}{T_2}. \label{eq:def_eff_and_ent}
\end{eqnarray}
The $2^\textrm{nd}$ law of thermodynamics guarantees that $\Delta S \geq 0$, with the equality satisfied only for a reversible process. It is easy to see that $\eta=\eta_C$ for a reversible process.

However, such exact reversible dynamics do not exist in the real world. Therefore, the attainability of the Carnot efficiency should be decided through a limiting process as follows. Define $\mathcal{A}$ as a set of parameters specifying a given heat engine. In this study, we will say ``the Carnot efficiency is reachable", if we can find some $\mathcal{A}$ satisfying
\begin{equation}
\eta_\textrm{C}-\eta<\epsilon \label{eq:Carnot_condition}
\end{equation}
for an arbitrary positive number $\epsilon$. From this viewpoint, we rewrite equation~(\ref{eq:def_eff_and_ent}) as
\begin{equation}
\eta_\textrm{C} -\eta =\frac{T_2 \Delta S}{|\mathcal{Q}_1|}. \label{eq:rewritten}
\end{equation}
Approaching a reversible process means the limit $\Delta S \rightarrow 0$ with finite $|\mathcal{Q}_1|$, which can be realized in a quasi-static process~\cite{Curzon}. In this limit, equation~(\ref{eq:Carnot_condition}) is satisfied, and thus, the Carnot efficiency is reachable with zero entropy production. On the other hand, for an irreversible process with finite $\Delta S >0$,
it has been widely accepted that the the Carnot efficiency can not be attainable and any irreversibility will reduce the engine efficiency.

However, there is another possibility for satisfying equation~(\ref{eq:Carnot_condition}). Imagine a heat engine
with non-zero entropy production $\Delta S $ and diverging heats $|\mathcal{Q}_1|$ and $|\mathcal{Q}_2|$ in some limit,
where leading diverging terms of $-|\mathcal{Q}_1|/T_1 +|\mathcal{Q}_2|/T_2$ cancel out each other in equation~(\ref{eq:def_eff_and_ent}).
%a limiting process $\Delta S/|\mathcal{Q}_1| \rightarrow 0$ with non-zero entropy production, i.e., $\Delta S \gtrsim O(1)$.
In this case, $\Delta S/|\mathcal{Q}_1| \rightarrow 0$, so the efficiency will also approach $\eta_\textrm{C}$.
As no such concrete example has yet been discovered before, it has been commonly misunderstood that $\eta_\textrm{C}$ is only reachable in the reversible limit. In this work, we present such an example explicitly for the first time and show that the Carnot efficiency is indeed reachable in an irreversible process. Note that recently studied engines achieving $\eta_\textrm{c}$ at finite power~\cite{Benenti,Allahverdyan,Koning} belong to the reversible limit case $(\Delta S=0)$~\cite{finite}.
%Note that, in usual irreversible processes, $\Delta S$ and $\mathcal{Q}_1$ are of the same order, so $\Delta S/|\mathcal{Q}_1|$ does not vanish.

We revisit and study the well-known Feynman-Smoluchowski ratchet (FSR) ~\cite{Smoluchowski,Feynman}
in a setup proposed by Sekimoto~\cite{Sekimoto}. The average heat transfers and the extracted work
 are calculated explicitly in the usual low temperature (or high energy barrier)
limit. We find that $\Delta S$ diverges but much slower than diverging $|\mathcal{Q}_1|$,
so $\Delta S/|\mathcal{Q}_1| \rightarrow 0$ in this limit. Hence, the Carnot efficiency is reachable
in the highly irreversible limit. We also find another counterintuitive and surprising result that
the irreversibility does not always reduce but {\em enhance} the engine efficiency in this model.

\section*{Model of the Feynman-Smoluchowski Ratchet}

\begin{figure}
\centering
\includegraphics[width=0.5\linewidth]{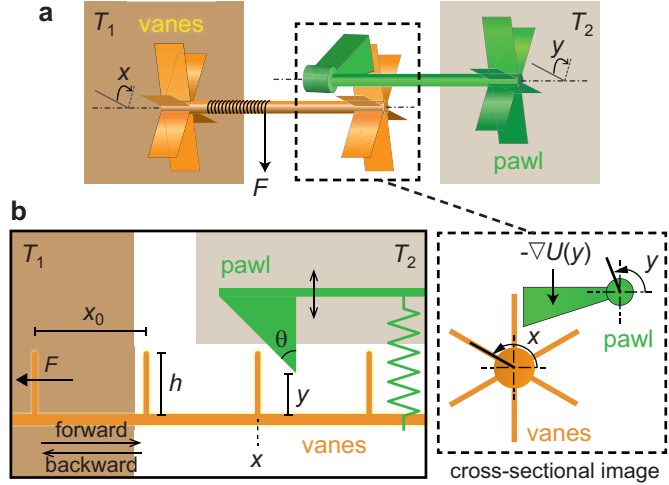}
\caption{\textbf{Schematic and model of the FSR.} (\textbf{a}) Schematic of the FSR. Vanes and a pawl are in contact with thermal baths of temperatures $T_1$ and $T_2$, respectively. $x$ and $y$ are the angles of the vanes and pawl, respectively. A constant load $F$ hangs from the axle of the vanes. The ratcheting interaction between the vanes and the pawl occurs outside of the baths, as illustrated in the boxed area.  Cross-sectional image: the ratcheting function is achieved by collision between the symmetric vanes and an angled pawl. $-\nabla U$ is a restoring force pulling down the pawl. (\textbf{b}) Schematic of the FSR model. One-dimensional vanes (the pawl) move only horizontally (vertically), and are in contact with thermal bath $T_1$ ($T_2$). $x$ is the position of one vane, $y$ is the height from the bottom of the vanes to the tip of the pawl, $F$ is a constant external force, $x_0$ is the distance between neighboring vanes, $h$ is height of a vane, and $\theta$ is angle of the pawl. The pawl is pulled down by a spring.
} \label{Fig:FSR}
\end{figure}

Figure~\ref{Fig:FSR}(a) shows a schematic of the FSR configuration, which consists of two components: vanes and a pawl. Both are in contact with different thermal baths of temperatures $T_1$ and $T_2$ ($T_1>T_2$), respectively, and their ratcheting interaction occurs outside of the baths. Ratcheting is achieved by interaction between the symmetric vanes and an angled pawl in our representation, while it takes place between a ratchet wheel with angled teeth and a simple pawl in the original FSR~\cite{Feynman}. However, both provide essentially the same rectifying function. Since only rotational motion is allowed, the dynamics of the vanes and the pawl can be described by their angles $x$ and $y$, respectively, which are stochastic variables due to thermal noise. Finally, a restoring force $-\nabla U$ pulls down the pawl and a constant load $F$ hangs on the axle of the vanes.

In this FSR setup, vanes are in contact only with a single heat bath at $T_1$ and heat flows from the hotter to the colder heat baths only through {\em mechanical collisions} between the vanes and the pawl~\cite{Sekimoto}.
Note that, in the original FSR~\cite{Feynman}, vanes are affected by two heat baths {\em simultaneously}  as illustrated in Supplementary Information (SI) Fig.~S1, where vanes can never be in equilibrium and thus heat should flow via vanes regardless of the mechanical interaction with the pawl~\cite{Parrondo,Magnasco}. In our setup, even in the presence of numerous mechanical
collisions, the vanes and the pawl can remain almost always in equilibrium with each bath, respectively, in the vanishing
limit of the mass ratio of the pawl and the vanes, which will be shown later. This is the key observation, which
makes it possible to reach the Carnot efficiency in the FSR.

The FSR as shown in Fig.~\ref{Fig:FSR}(a) is modeled as illustrated in Fig.~\ref{Fig:FSR}(b). For simplicity, we assume that the one-dimensional vanes move only horizontally and the pawl moves only vertically~\cite{2d}. They are in contact with thermal baths $T_1$ and $T_2$, respectively, and the ratcheting interaction occurs outside of the baths. $x$ is the position of one vane and $y$ is the height from the bottom of vanes to the tip of the pawl. Since the pawl cannot penetrate the bottom, $y\geq 0$. The vanes and the pawl are pulled by the constant external force $F$ and the harmonic force $-ky$, respectively. Here, the direction against $F$ is defined as `forward'. $x_0$ is the distance between neighboring vanes, $h$ is the height of a vane, and $\theta$ is the angle of the pawl. Then, the corresponding Langevin equation can be written as
\begin{eqnarray}
&\textrm{vane:}& v = \dot{x},~
m\dot{v} = -F - g_\textrm{v}(x,y) - \gamma_1 v +\xi_1, \label{eq:Langevin_vane} \\
&\textrm{pawl:}& u = \dot{y},~
m_\textrm{p} \dot{u} = g_\textrm{p}(x,y) - ky - \gamma_2 u +\xi_2~(y\geq 0),~~~
\label{eq:Langevin_pawl}
\end{eqnarray}
where $m$ and $m_\textrm{p}$ are the masses of the vanes and the pawl respectively, $\gamma_i$ is the damping coefficient of heat bath $i$, and $\xi_i(t)$ is the Gaussian noise of heat bath $i$ at time $t$ satisfying $\langle \xi_i(t) \xi_j(t')\rangle =2\gamma_i T_i \delta_{ij} \delta(t-t')$ (the Boltzmann constant is set to $k_B=1$).
$g_\textrm{v}(x,y)$ and $g_\textrm{p}(x,y)$ denote the forces exerted to the vanes and the pawl, respectively, through elastic collisions between a vane and the pawl. The detailed forms of the forces are given in SI Fig.~S2.

\begin{figure}
\centering
\includegraphics[width=0.5\linewidth]{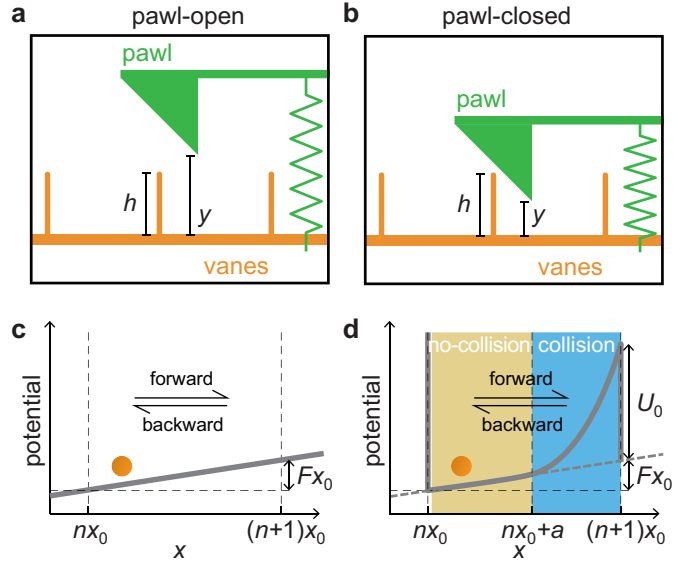}
\caption{\textbf{Schematics of the pawl-open and pawl-closed states.} (\textbf{a}) Pawl-open state ($y>h$). (\textbf{b}) Pawl-closed state ($y\leq h$). (\textbf{c}) Potential in the pawl-open state. Only a linear potential with slope $F$ is felt by the vanes. (\textbf{d}) Potential in the pawl-closed state when $n x_0\leq x<(n+1)x_0$ ($n$ is an integer). An infinite potential wall at $n x_0$ prevents a backward hop. There are two regions: the no-collision ($n x_0< x<n x_0 +a$) and collision regions ($n x_0+a \leq x<(n+1)x_0 $), depending on whether a collision between the vanes and the pawl takes place. $U_0$ is the potential energy of the pawl at $y=h$.}
\label{Fig:two_state}
\end{figure}

We define two states in this model: the pawl-open and pawl-closed states as shown in Fig.~\ref{Fig:two_state} (a) and (b), respectively. In the pawl-open state ($y>h$), both forward and backward hopping movements of the vanes are possible. Here, one hop denotes movement of $x$ from $n x_0<x<(n+1)x_0$ ($n$ is an integer) to $n^\prime x_0<x<(n^\prime+1)x_0$ ($n^\prime=n\pm 1$). Since $g_\textrm{v}(x,y)=0$ in this state, only a linear potential with slope $F$ is felt by the vanes, as shown in Fig.~\ref{Fig:two_state}(c). As there is no interaction between the vanes and the pawl in this state,
no energy is transferred from the vanes to the pawl.

In the pawl-closed state ($y\leq h$), the pawl completely forbids a backward hop of the vanes at $x=n x_0 $. This blockage is felt by the vanes as an infinite potential barrier located at $x=n x_0 $, as illustrated in Fig.~\ref{Fig:two_state}(d). Note that no energy is transferred to the pawl by this blocking collision because a horizontal force does not induce any (vertical) $y$ displacement~\cite{2d2}. For $n x_0<x<nx_0 +a $ ($a\equiv x_0-h$), the vanes feels only a linear potential of slope $F$ without any collision, i.e., $g_\textrm{v}(x,y)=g_\textrm{p}(x,y)=0$.  For $n x_0 +a\leq x<(n+1)x_0 $ ({\em collision} region),
the vane and the pawl collide with each other and some energy is transferred from the vanes to the pawl, which
is eventually dissipated as heat $Q_\textrm{col}$ into the heat bath $2$.
Once in a while when high enough thermal energy is supplied to the vanes from the heat bath 1, the vane can go over $(n+1)x_0$ by lifting the pawl
up to $y=h$ by the collision. In this case, the vanes should overcome an energy barrier of height $U_0+Fx_0$ with $U_0=kh^2/2$. In this one-step forward hopping process, the energy delivered to the pawl from the vanes is $U_0$, which is dissipated as heat $Q_\textrm{hop}$ in the heat bath $2$, i.e., $\Delta Q_\textrm{hop}=U_0$ per one hopping~\cite{overcoming}.
Then, $Q_2 = Q_\textrm{col}+Q_\textrm{hop}$, where $Q_2$ is heat dissipation into the heat bath $2$.
%because there is no other contribution for $Q_2$.
For later discussion we define the average time for one forward hop as $\tau_\textrm{hop}$.

\section*{High energy barrier or low temperature limit}

We now consider the high energy barrier (low temperature) limit:
\begin{equation}
T_2 <T_1 \ll U_0, Fx_0~ \label{eq:limit}
%1 \ll \frac{Fx_0}{T_1}, \frac{U_0}{T_1} \ll \frac{U_0}{T_2} \quad (T_2<T_1).
\end{equation}
with an additional condition $\eta_C U_0/T_2 \gg 1$ for later convenience.
For large $U_0/T_2$, the FSR will almost always be in the pawl-closed state due to huge energy barriers against
  thermal fluctuations. Along with large $Fx_0/T_1$, the vanes will rarely reach the collision region against a very steep energy hill. Therefore, in the above limit, the vanes will spend most of their time in the no-collision region ($nx_0< x<nx_0+a$). Then, equation~(\ref{eq:Langevin_vane}), the dynamics of vanes, can be approximately written as
\begin{equation}
\textrm{vane:}~~ v = \dot{x},~
m\dot{v} = -F - \gamma_1 v +\xi_1~(x\geq nx_0),~~~ \label{eq:Langevin_vane_equil}
\end{equation}
with an infinite energy barrier at $x=nx_0$. Similarly, equation~(\ref{eq:Langevin_pawl}) can be practically written as
\begin{equation}
\textrm{pawl:}~~ u = \dot{y},~
m_\textrm{p} \dot{u} = - ky - \gamma_2 u +\xi_2~(y\geq 0),~~~
\label{eq:Langevin_pawl_equil}
\end{equation}
with an infinite energy barrier at $y=0$. This implies that the steady-state probability distributions of the vanes and the pawl are almost the same as the equilibrium distributions of the Langevin equations~(\ref{eq:Langevin_vane_equil}) and (\ref{eq:Langevin_pawl_equil}), respectively, in the high energy barrier limit, which will be confirmed numerically later. Hence, the probabilities for the pawl-open and pawl-closed states, $p_\textrm{o}$ and $p_\textrm{c}$, become
\begin{eqnarray}
p_\textrm{o} &\approx& \int_h^\infty dy \sqrt{\frac{2k}{ \pi T_2}} e^{-\frac{k y^2}{2T_2}}
\approx \sqrt{\frac{T_2}{ \pi U_0}}e^{-\frac{U_0}{ T_2}}~,~~ \nonumber \\
p_\textrm{c} &=& 1-p_\textrm{o} \approx 1, \label{eq:state_prob}
\end{eqnarray}
respectively. Note that all higher-order corrections are exponentially small in $U_0/T_2$.

In this limit, we estimate the power $\langle \dot{W} \rangle_s $ and the heat dissipation rate into the heat bath $2$ $\langle \dot{Q}_2 \rangle_s $, where $\langle \cdots \rangle_s$ denotes the steady-state average. These can be written as
\begin{eqnarray}
\langle \dot{W} \rangle_s &=& (r_\textrm{f}-r_\textrm{b}) Fx_0,\nonumber \\
\langle \dot{Q}_2 \rangle_s &=& \langle \dot{Q}_\textrm{col} \rangle_s+\langle \dot{Q}_\textrm{hop} \rangle_s , \label{eq:energy_rate}
\end{eqnarray}
where $r_\textrm{f}$ and $r_\textrm{b}$ are the rates of forward and backward hopping, respectively. The power $\langle \dot{W} \rangle_s $ is the work rate of lifting the load hanging from the axle of the vanes. The heat dissipation can be separated into two terms based on collisions and hopping, as discussed before.

First, consider the rates in the pawl-closed state. Since the vanes are almost always in equilibrium at $T_1$, the rate of forward hopping that overcomes an energy barrier of height $U_0+Fx_0$ can be estimated from the Arrhenius rate equation as $r_\textrm{f,c} \approx N_\textrm{c} p_\textrm{c} e^{-(Fx_0+U_0)/T_1}$ where $N_\textrm{c}$ is the hopping-attempt frequency~\cite{Hanggi_rev,Arrhenius} of the pawl-closed state. The backward hopping rate $r_\textrm{b,c}$ is simply zero in this case. In the pawl-open state, both forward and backward hops are possible, but the forward hopping rate $r_\textrm{f,o}$ is exponentially smaller ($\sim e^{-Fx_0/T_1}$) than the backward hopping rate $r_\textrm{b,o}$ for large $Fx_0/T_1$. So, the backward hopping rate will be almost identical to the hopping-attempt frequency, i.e., $r_\textrm{b,o} \approx N_\textrm{o}p_\textrm{o} $ with exponentially small corrections.
Note that $N_\textrm{o}$ and $N_\textrm{c}$ can differ, but this difference will not be very large.
The vanes spend most of time in the pawl-closed state and become fully relaxed. As the pawl opens for a very short period
($\sim \tau_\textrm{hop} p_\textrm{o} \sim T_2/U_0$), we expect that the vane statistics does not deviate significantly from the fully relaxed one. Therefore, $N_\textrm{o} / N_\textrm{c} $ can be reasonably assumed to be a constant of $O(1)$. Then, we have
\begin{eqnarray}
&&r_\textrm{f}= r_\textrm{f,o} + r_\textrm{f,c} \approx r_\textrm{f,c} \approx N_\textrm{c} e^{-\frac{U_0+Fx_0}{T_1}}, \nonumber \\
&&r_\textrm{b}= r_\textrm{b,o} + r_\textrm{b,c} = r_\textrm{b,o} \approx N_\textrm{o}\sqrt{\frac{T_2}{ \pi U_0}} e^{-\frac{U_0}{T_2}}, \label{eq:rate}
\end{eqnarray}
where $r_\textrm{f,o}$ is ignored since $r_\textrm{f,o}/r_\textrm{f,c}\sim (T_2/U_0)^{1/2} e^{-\eta_C U_0/T_2}$ .

Now consider $\langle \dot{Q}_2 \rangle_s$. Backward hopping occurs only when the system is in the pawl-open state. Thus, there is no heat dissipation into the heat reservoir 2, associated with backward hopping~\cite{back}.
Hence, $\langle \dot{Q}_\textrm{hop} \rangle_s = r_\textrm{f,c} U_0$, since $\Delta Q_\textrm{hop} = U_0$ per one forward hopping. It is not trivial to estimate $\langle \dot{Q}_\textrm{col} \rangle_s$, which originates from the energy transfer due to numerous collisions between a vane and the pawl in the collision region of the pawl-closed state, before finally going over the hopping energy barrier. In our elastic collision model (SI Fig.~S2), it is easy to show that the transferred energy per collision is linearly proportional to the mass ratio $m_\textrm{p}/m$ for small $m_\textrm{p}/m$ (see SI Sec.~1).
%This is consistent with a simple intuition that the kinetic energy change of a heavy
%object should be negligible through collision with a very light object.
On the other hand, one may expect that the collision frequency diverges in the limit of $m_p/m\rightarrow 0$,
though it is difficult to derive the average rate of total energy transfer analytically in terms of the mass ratio even without thermal noises.
Nevertheless, numerical simulations confirm that
$\langle \dot{Q}_\textrm{col} \rangle_s$ indeed vanishes in this limit~\cite{fsr,CN}  as
\begin{equation}
\langle \dot{Q}_\textrm{col} \rangle_s \sim N_cp_c \left(m_\textrm{p}/m\right)^{\omega} \left(T_1-T_2\right), \label{eq:Q_int}
\end{equation}
with $\omega=0.27(3)$. Details of our simulation results will be shown later.
It is crucial to notice that $\langle \dot{Q}_\textrm{col} \rangle_s$ can be made  arbitrarily smaller than $\langle \dot{Q}_\textrm{hop} \rangle_s$, i.e. $\langle \dot{Q}_\textrm{col} \rangle_s \ll \langle \dot{Q}_\textrm{hop} \rangle_s$, by taking an appropriately small value of the mass ratio $m_\textrm{p}/m$ as
\begin{equation}
%\frac{m_\textrm{p}}{m} \ll  \left[\frac{e^{-(U_0+Fx_0)/T_1} U_0}{\eta_C T_1}\right]^{1/\omega}~.
m_\textrm{p}/m \ll  \left[{e^{-(U_0+Fx_0)/T_1} U_0}/({\eta_C T_1})\right]^{1/\omega}~.
\label{eq:mass_ratio}
\end{equation}%~\cite{frequency}.
Therefore, in this small mass ratio limit, we get
\begin{equation}
\langle \dot{Q}_2 \rangle_s \approx \langle \dot{Q}_\textrm{hop}\rangle_s \approx r_\textrm{f} U_0, \label{eq:Q_2}
\end{equation}

%%%%%%%%%%%%%%%%%%%%%%%%%%%%%%%%%%%%%%%%%%%%%%%%%%%%%%%

Using equations~(\ref{eq:energy_rate}), (\ref{eq:rate}), and (\ref{eq:Q_2}), we calculate the efficiency and entropy production in both the high energy barrier and the small mass ratio limits.
First, the efficiency is given by
\begin{eqnarray}
\eta = \frac{\langle \dot{W} \rangle_s  }{\langle \dot{W} \rangle_s  +\langle \dot{Q}_2 \rangle_s }
 \approx \frac{(r_\textrm{f}-r_\textrm{b}) Fx_0}{(r_\textrm{f}-r_\textrm{b}) Fx_0+r_\textrm{f}U_0}
%  = \frac{ g(z)}{1+g(z)} \quad \textrm{with} ~~ z=\frac{F x_0}{T_1}\left(\frac{T_2}{U_0 \eta_C}\right)
, \label{eq:eta}
\end{eqnarray}
For convenience, this can be rewritten in terms of a dimensionless external load $z$ as
\begin{equation}
\eta (z) = \frac{ \eta_C g(z)}{1-\eta_C [1-g(z)]} \qquad \textrm{with} ~~ z=\frac{F x_0}{T_1}\left(\frac{T_2}{\eta_C U_0 }\right)
, \label{eq:eta1}
\end{equation}
where
\begin{equation}
g(z)= z \left[1-\frac{r_\textrm{b}}{r_\textrm{f}}\right]= z \left[1-\beta e^{-\frac{\eta_C U_0}{T_2} (1-z)}\right] \qquad \textrm{with} ~~ \beta=\frac{N_o}{N_c}\sqrt{\frac{T_2}{\pi U_0}}~.
%\alpha = \frac{U_0}{T_2}-\frac{U_0+Fx_0}{T_1}~~\text{and}~~\beta=\frac{N_\textrm{o}}{N_\textrm{c}} \sqrt{\frac{T_2}{\pi U_0}}.
\end{equation}
To be a useful heat engine (positive work extraction against the load),  $g(z)$ should be larger than zero. Moreover,
since $Fx_0\ge 0$, we have the condition for $z$ as
\begin{equation}
0 \le z \le z^s \quad \textrm{with}~~ z^s=1-\frac{T_2}{\eta_c U_0} \ln \beta
\approx 1+\frac{T_2}{2\eta_c U_0}\ln \left( \frac{U_0}{T_2}\right) +O\left( \frac{T_2}{\eta_C U_0} \right)~,
\label{range}
\end{equation}
where the average speed of the vanes is zero at $z=z^s$ ({\em stalling} point: $r_\textrm{f}=r_\textrm{b}$ and $\langle \dot{W} \rangle_s  =0$).
%where
%\begin{equation}
%\alpha_m=U_0\left( \frac{1}{T_2}- \frac{1}{T_1}\right)=\frac{\eta_C U_0}{T_2}.
%\end{equation}

For fixed $U_0/T_1$ and $U_0/T_2$, we find the maximum efficiency $\eta^m$ by varying the external load $z$ in the range of equation~(\ref{range}):
$d\eta(z)/dz|_{z=z^m} =0$. The result is
\begin{equation}
z^m \approx 1- \frac{T_2}{\eta_C U_0 } \ln \left(\frac{\beta \eta_C U_0}{T_2}\right) \approx
1- \frac{T_2}{2 \eta_C U_0 } \ln \left(\frac{\eta_C U_0}{T_2}\right)
+O\left( \frac{T_2}{\eta_C U_0} \right)
, \label{eq:alpha}
\end{equation}
which is well inside of the range of equation~(\ref{range}).
Plugging this into equation~(\ref{eq:eta1}), it is easy to see
\begin{equation}
\eta^m =\eta(z^m) \approx \eta_C -
 \frac{(1-\eta_C)T_2}{U_0}\ln \left(\frac{\beta\eta_C U_0}{T_2}\right)
+O\left( \frac{T_2}{U_0} \right). \label{eq:eff_bound}
\end{equation}
This clearly shows that the Carnot efficiency $\eta_C$ can be reached in the high energy barrier limit.

Interestingly, the maximum efficiency is obtained {\em not} at the stalling point (usual in the reversible engine),
but $z^m$ and $z^s$ approach to $z=1$ from the above and the below, respectively, in the high energy barrier limit.
Furthermore, the backward hopping is negligible at $z=z^m$ as
$r_\textrm{b}/r_\textrm{f}\propto T_2/(\eta_C U_0)$ and the average power is obtained as
\begin{equation}
\langle \dot{W} \rangle_s^m\approx r_\textrm{f} F x_0|_{z=z^m} \approx r_\textrm{f}(z^m) T_1\left[ \frac{\eta_C U_0}{T_2}-\ln \left(\frac{\beta\eta_C U_0}{T_2}\right)\right]
\qquad \textrm{with} ~~r_\textrm{f}(z^m)\approx N_c e^{-\frac{U_0}{T_2} +\ln \left(\frac{\beta\eta_C U_0}{T_2}\right)}~.
\end{equation}
%with
%\begin{equation}
%r_\textrm{f}\approx N_c e^{-\frac{U_0}{T_2} +\frac{1}{2}\ln (\eta_C U_0/T_2)}~.
%\end{equation}
%%%%%%%%%%%%%%%%%%%%%%%%%%%%%%%%%%%%%%%%%%%%%
The average time for one forward hop should be given as the inverse of the forward hopping rate as $\tau_\textrm{hop}\approx 1/r_\textrm{f}$, which diverges exponentially with $U_0/T_2$. This implies that our FSR operates very slowly with a moderate value of $N_c$, similar to an ordinary Carnot engine operating in a quasi-static way. The power generation is also vanishingly small due to the exponentially diverging hopping period, but the work extraction is very large (proportional to $\eta_C U_0$) in one hopping duration, in contrast to the finite work extraction in the ordinary Carnot engine.

The steady-state entropy production (EP) rate $\langle \dot{S}\rangle_s$ can be also evaluated from equation~(\ref{eq:def_eff_and_ent}), with the average heat transfer rate from heat bath 1, $\langle \dot{Q}_1\rangle_s=\langle \dot{Q}_2\rangle_s+\langle \dot{W} \rangle_s$, given as
\begin{equation}
\langle\dot{S}\rangle_s = -\frac{\langle \dot{Q}_1\rangle_s}{T_1} + \frac{\langle \dot{Q}_2 \rangle_s}{T_2}
= r_\textrm{f}(z) \frac{\eta_C U_0}{T_2} \left[1- g(z)\right]
\qquad \textrm{with} ~~r_\textrm{f}(z)= N_c e^{-\frac{U_0}{T_1} -\frac{\eta_C U_0 }{T_2}z}~.
\label{eq:ep1}
\end{equation}
The EP rate at the maximum efficiency point ($z=z^m$) is
\begin{equation}
\langle\dot{S}\rangle_s^m
\approx  r_\textrm{f}(z^m) \ln \left(\frac{\beta \eta_C U_0}{T_2}\right) \approx \frac{1}{2} r_\textrm{f}(z^m) \ln \left(\frac{\eta_C U_0}{T_2}\right) ~,
\end{equation}
where the most dominant terms linearly proportional
to $U_0/T_2$ cancel out each other.
This rate is again vanishingly small, but the entropy production during one hopping period $\Delta S$ becomes
\begin{equation}
\Delta S = \tau_\textrm{hop} \langle\dot{S}\rangle_s \approx \frac{1}{2}\ln \left(\frac{\eta_C U_0}{T_2}\right),
\end{equation}
which can be very large. Therefore, the FSR operates definitely in a strongly {\em irreversible} process, while retaining the Carnot efficiency in the high energy barrier limit. In terms of equation~(\ref{eq:rewritten}), both $\Delta S$ and $|\mathcal{Q}_1|$ during one hopping period diverge, but in a different manner to $\Delta S\propto \ln (\eta_C U_0/T_2)$ and $|\mathcal{Q}_1|/T_1 \propto U_0/T_2$, thus its ratio approaches zero in the $U_0/T_2\rightarrow \infty$ limit (see also equation~(\ref{eq:eff_bound})), which is in sharp contrast to the conventional reversible Carnot engine~\cite{rev}.

\begin{figure}
\centering
\includegraphics[width=0.5\linewidth]{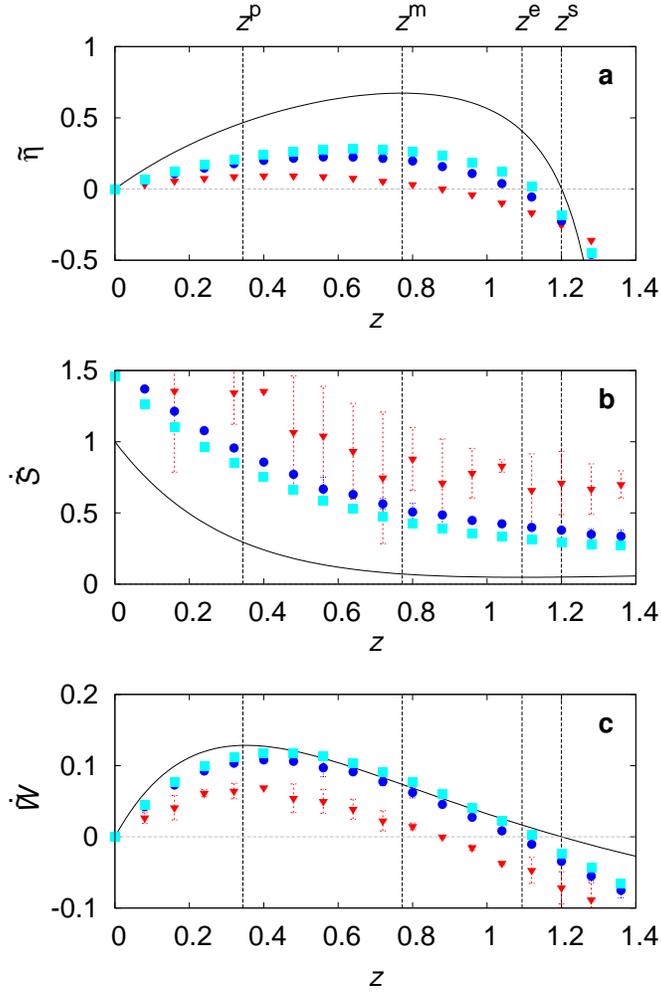}
\caption{\textbf{Efficiency, entropy production (EP) rate, and power.}
(\textbf{a}) The properly scaled dimensionless efficiency ${\tilde \eta}=\eta/\eta_C$,
(\textbf{b}) the EP rate
$\dot{\tilde S}=\langle \dot{S}\rangle_s/[N_c e^{-U_0/T_1} (\eta_C U_0/T_2)]$, and (\textbf{c})
the power  $\dot{\tilde W}=\langle \dot{W}\rangle_s/[N_c e^{-U_0/T_1} (\eta_C U_0/T_2)T_1]$
are plotted against the dimensionless external load $z=Fx_0T_2/(\eta_C U_0 T_1)$ (solid lines).
Four special points are denoted as $z^p$ (maximum power), $z^m$ (maximum efficiency),
$z^e$ (minimum EP rate), and $z^s$ (stalling: $r_\textrm{f}=r_\textrm{b}$).
We take $T_1=2$, $T_2=1$, $U_0=5$, $x_0=2$, $N_c=0.045$, $N_0/N_c=2.4$
and vary $F$ from 0 to $3.5$.
%In the high energy barrier limit ($\eta_C U_0/T_2 \rightarrow \infty$), $z^p$ approaches $z=0$, and
%all the other three points ($z^m$, $z^e$, $z^s$) approaches $z=1$.
In the region of $z^m <z<z^e$, the larger the irreversibility $\langle \dot{S}\rangle_s$, the higher the efficiency $\eta$.
Simulation data averaged over 10 steady states up to the simulation time $\tau=2\times 10^{11}$
are denoted by symbols for various values of the mass ratio:
$m_\textrm{p}/m=10^{-1} (\triangledown)$, $10^{-2} (\bigcirc)$, and $10^{-3} (\square)$.
} \label{Fig:EEP}
\end{figure}

It is also interesting to study the behavior of the EP rate as a function of the external load $z$.
In a similar way to the above, we find that the EP rate is minimized at $z=z^e$ as
\begin{equation}
z^e\approx 1+\frac{T_2}{\eta_C U_0}\left[1-e\beta+O(\beta^2)\right]~,
\label{eq:zs}
\end{equation}
which is again inside of the range of equation~(\ref{range}), but larger than the maximum efficiency point $z^m$
($z^m <1<z^e<z^s$). This point $z^e$ also approaches to $z=1$ as well as the other two points
in the high energy barrier limit, but in a different fashion.
The efficiency $\eta$ and the EP rate $\langle\dot{S}\rangle_s$ are plotted against the external load $z$ in Fig.~\ref{Fig:EEP}. The solid lines in (a) and (b) are drawn by equations~\eqref{eq:eta1} and \eqref{eq:ep1} with $U_0/T_2=5$.
We can see that efficiency increases rapidly when the EP rate increases slightly in the region of $z^m<z<z^e$. This shows that {\em increasing irreversibility can drastically enhance the efficiency
in a highly irreversible process}, which is quite surprising and against the conventional wisdom. Note that the EP rate does not go to zero even at the stalling point ($z=z^s$) for large but finite $\eta_C U_0/T_2$. The values of the EP rate and the power at this EP minimum point are calculated
as
\begin{equation}
\langle\dot{S}\rangle_s^e \approx r_\textrm{f}(z^e) e\beta \frac{\eta_C  U_0}{ T_2}
\sim r_\textrm{f}(z^e) \left(\frac{\eta_C  U_0}{ T_2}\right)^{1/2}
\quad \textrm{and} \quad \langle\dot{W}\rangle_s^e \approx r_\textrm{f}(z^e)
T_1\frac{\eta_C U_0}{T_2}\left( 1-e\beta\right)
\quad \textrm{with} ~~r_\textrm{f}(z^e)= N_c e^{-\frac{U_0}{T_2} -1}~.
\end{equation}

Finally, we also investigate when the maximum power is achieved. The results are
\begin{equation}
 z^p\approx \frac{T_2}{\eta_C U_0} \left( 1-e\beta e^{-\frac{\eta_C U_0}{T_2}}\right)~,
\quad \langle\dot{S}\rangle_s^p \approx  r_\textrm{f}(z^p) \frac{\eta_C  U_0}{ T_2}~,
\quad \textrm{and}  \quad \langle\dot{W}\rangle_s^p \approx r_\textrm{f}(z^p) T_1~
\quad \textrm{with} ~~r_\textrm{f}(z^p)= N_c e^{-\frac{U_0}{T_1} -1} .
\label{eq:zp}
\end{equation}
Note that the maximum power is generated at a very small load $z^p$ for large $\eta_C U_0/T_2$. The power,
$\langle \dot{W}\rangle_s=  r_\textrm{f}(z) T_1  g(z) \eta_C U_0/T_2$, is also
plotted in Fig.~\ref{Fig:EEP} (c). The efficiency at the maximum power point can be obtained as
$\eta(z^p)\approx T_2/[(1-\eta_C)U_0]$, which vanishes in the high energy barrier limit.
As our FSR cannot be described by the linear response theory (highly irreversible), the efficiency
at the maximum power does not take any universal form, discussed in recent literatures~\cite{CA,ELB}.

\section*{Numerical Evidences}

We performed a numerical simulation to check the validity of our theory. In this simulation, we numerically integrated the Langevin equations~(\ref{eq:Langevin_vane}) and (\ref{eq:Langevin_pawl}), using a second-order integrator~\cite{Ciccotti}. To implement the interaction forces $g_\textrm{v}(x,y)$ and $g_\textrm{p}(x,y)$, we assumed that collisions between a vane and the pawl are elastic and instantaneous (see SI Fig.~S2). For convenience, we used dimensionless variables by rescaling time, length, and energy in units of $\gamma_0/k_0$, $\sqrt{T_2/k_0}$, and $T_2$, respectively. Here $\gamma_0$ and $k_0$ are constants with dimensions of the damping coefficient and the spring constant, respectively. Heat can be calculated as $Q_1=\int_0^\tau dt ~v \circ (-\gamma_1 v+ \xi_1)$ and $Q_2=-\int_0^\tau dt ~u \circ (-\gamma_2 u+ \xi_2)$ during $\tau$, where $\circ$ denotes the Stratonovich integral~\cite{Sekimoto,gardiner}. Then, the heat dissipation rates $\langle \dot{Q}_i \rangle_s \equiv Q_i/\tau$ in a steady state. For convenience, we take $T_1 =2$, $T_2 =1$, $m=10$, $\gamma_1=\gamma_2=1$, and $\theta=45^\circ$.

\begin{figure}
\centering
\includegraphics[width=0.5\linewidth]{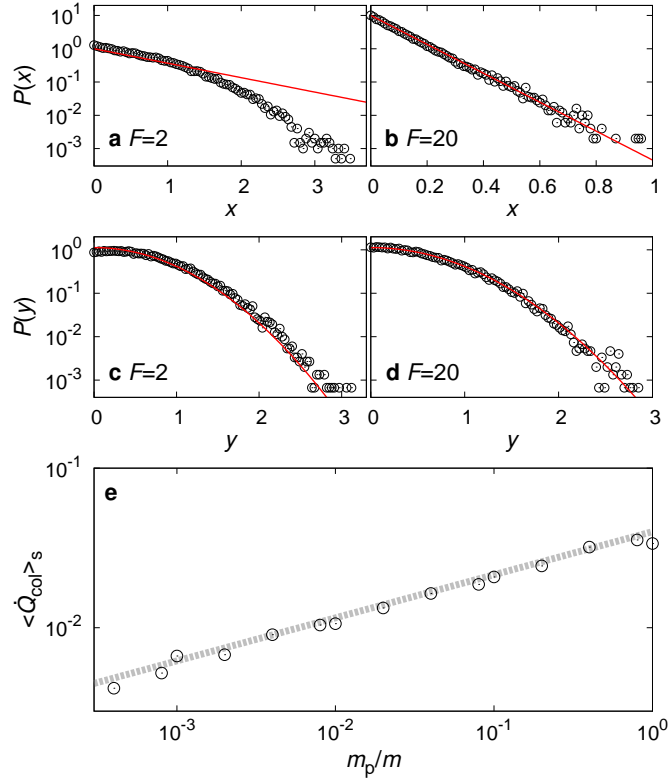}
\caption{\textbf{Numerical results.} (\textbf{a}) and (\textbf{b}) show the probability distributions of $x$ for $F=2$ and $F=20$, respectively. (\textbf{c}) and (\textbf{d}) show the probability distributions of $y$ for $F=2$ and $F=20$, respectively. Solid curves denote equilibrium distributions of equations~(\ref{eq:Langevin_vane_equil}) and (\ref{eq:Langevin_pawl_equil}), respectively. (\textbf{e}) Log-log plot of $\langle \dot{Q}_\textrm{col} \rangle_s$ versus $m_\textrm{p}/m$. The dashed line is a guide line with slope $0.27$.} \label{Fig:numerical}
\end{figure}

We first check whether the vanes and the pawl are almost always in equilibrium as described by equations~(\ref{eq:Langevin_vane_equil}) and (\ref{eq:Langevin_pawl_equil}), respectively, in the high energy barrier limit. We set $m_\textrm{p}=0.1$, $x_0=11$, $k=2$, and $h=10$. Since $U_0=kh^2/2=100$ is much larger than thermal energies, $T_1$ and $T_2$, no forward or backward hops can occur within our simulation time ($\tau = 7.5\times 10^7$), so $x$ remains between $0$ and $x_0$ at $n=0$. Figures~\ref{Fig:numerical} (a) and (c) show the probability distributions of $x$ and $y$ for $F=2$, respectively. They show clear deviations from the equilibrium distributions (solid lines), due to energy transfer via numerous collisions between the vanes and the pawl for small $F$. However, for $F=20$, we can see perfect agreement in Figs.~\ref{Fig:numerical} (b) and (d).

We also checked the validity of equation~(\ref{eq:Q_int}). For better statistics of numerical data, we use
a lighter  load (small $F$) to facilitate more collisions. We set $F=1$, $x_0=2$, $a=1$, $k=100$ and $h=1$. Since $U_0=50$ is still large, no hopping occurs  within our simulation time ($\tau = 2 \times 10^9$) and the heat dissipation into the heat reservoir 2 is solely from energy transfer via collisions, i.e., $\langle \dot{Q}_2 \rangle_s =\langle \dot{Q}_\textrm{col} \rangle_s$. Figure~\ref{Fig:numerical}(e) shows the log-log plot for $\langle \dot{Q}_\textrm{col} \rangle_s$ versus $m_\textrm{p}/m$, which shows a power-law scaling of $\langle \dot{Q}_\textrm{col} \rangle_s$ with the exponent 
$\omega=0.27(3)$, which confirms equation~(\ref{eq:Q_int}).

It is practically infeasible to measure the efficiency numerically for large $U_0/T_2$ in our simulation time ($\tau=2 \times 10^{11}$), because $\tau_\textrm{hop}$ grows exponentially with $U_0/T_2$. Instead, we obtained the data at a rather small value of $U_0/T_2=5$ by varying $F$ from 0 to 3.5 with $x_0=2$, which are presented in Fig.~\ref{Fig:EEP} for several different values of the mass ratio
($m_\textrm{p}/m=10^{-1}, 10^{-2}, 10^{-3}$).
Even in this case, it is remarkable to see that all data sets for the efficiency, the EP rate, and the power show general features quite consistent
with the analytic predictions such as the locations of the maximum efficiency point ($z^m$), the minimum EP rate point
($z^e$), and the maximum power point ($z^p$). The proper criterion for the small mass ratio limit given by equation
(\ref{eq:mass_ratio}) is $m_\textrm{p}/m \ll 3.5\times 10^{-6}$ near $z=1$. So, it is not surprising to see that
the EP rate with $m_\textrm{p}/m=10^{-3}$ is quite higher than the analytic prediction in Fig.~\ref{Fig:EEP} (b), due to non-negligible heat dissipation due to collisions, $\langle \dot{Q}_\textrm{col}\rangle_s$. Accordingly, the efficiency is also lower in Fig.~\ref{Fig:EEP} (a), which is expected to approach the
analytically predicted line with $m_\textrm{p}/m \approx 10^{-6}$. The power data (only depending on the hopping frequencies
$r_\textrm{f}$ and $r_\textrm{b}$) are in an excellent agreement with
the theoretical prediction already with $m_\textrm{p}/m =10^{-3}$.
Most importantly, our simulation data with
a finite mass ratio value and a moderate value of $U_0/T_2$ still show that
the larger the irreversibility the higher the efficiency in some region near the maximum efficiency
($z^m < z < z^e$). This suggests that this counterintuitive prediction can be rather easily observed in realistic situations
by experiments or simulations in highly irreversible environments. In a small system such as a kinesin molecular motor inside
a biological cell, the large attempt frequency $N_c$ makes  hoppings very frequent with $\tau_\textrm{hop}\approx 10^{-2} sec$
for typical energy barriers $U_0/T_2\approx 8$, and $Fx_0/T_2\approx 12$~\cite{kinesin}. This may serve as one of many possible examples to investigate systematically the relation between the heat dissipation and the efficiency.

%\section*{Experimental }

\section*{Summary and Discussion}

In summary, we have described a heat engine that can operate with the Carnot efficiency in an irreversible process. It has a vanishing power and a vanishing entropy production (EP) rate. However, during one cycle (forward hop), the extracted work, the heat currents, and the entropy production all diverge in the Carnot efficiency limit, which makes the process fully irreversible, in contrast to the conventional reversible Carnot engine. The key observation is that the EP divergence is weaker than the divergence of the heat currents to achieve the Carnot efficiency. Our result is consistent with the recent rigorous bound claiming that power should go to zero when the efficiency approaches the Carnot efficiency~\cite{Tasaki}.

We also find another surprising result that the irreversibility can enhance the engine efficiency.
Until now, there has been a conventional misbelief that the irreversibility inevitably reduces the efficiency, and thus decreasing the irreversibility is the only way to get a higher efficiency. Thus,
our finding opens a new possibility to develop a novel design of thermodynamic engines, especially for microscopic ones actively studied recently~\cite{Martinez, Blickle, Koski}, with a high efficiency in highly irreversible processes.

Our results are based on the careful setup of the FSR (mechanical collisions between the vanes and the pawl
outside of both heat baths) and two key limits: the high energy barrier limit and the small mass ratio limit.
In case of the original FSR setup where the vanes are in contact with both baths simultaneously (SI Fig.~S1), 
it is impossible to reach the Carnot efficiency due to the existence of uncontrollable irreversible heat currents.
The high energy barrier limit ensures that the vanes and the pawl are almost always in equilibrium with each bath
and the small mass limit controls the irreversible heat current arising from numerous collisions without a hop to be vanishingly small. Numerical simulations support our results very well. In particular, the interesting possibility that the larger the irreversibility the higher the efficiency can be observed by experiments or by numerical simulations in realistic situations (small systems) quite far from the both limits. More explicit applications in nano and bio systems may be well expected.

\section*{Acknowledgements}

This research was supported by the NRF grant No.~2011-35B-C00014 (JSL).
We thank Changbong Hyun for many useful discussions, and also Hyun-Myung Chun and Jae Dong Noh for
sending their unpublished results.

\section*{Author contributions statement}

J.S.L designed the study and performed simulations. J.S.L. and H.P. discussed and wrote the manuscript together.

\section*{Additional information}

\textbf{Competing financial interests} The authors declare no competing financial interests.


\begin{thebibliography}{99}
\bibitem{finite} Equation~\eqref{eq:rewritten} can be rewritten as $\eta_\textrm{C}/\eta -1 = T_2  \Delta S / W$
with the extracted work $W=| \mathcal{Q}_1 |-| \mathcal{Q}_2 |$. With finite $W$ (nonzero power with finite duration time), it is obvious that $\Delta S$ must be zero in order to attain the Carnot efficiency. Therefore, all engines having the Carnot efficiency with a finite power, if exists, should be a reversible engine.
\bibitem{2d} In this study, we consider the situation where the vanes and the pawl move along one-dimensional frictionless rails with infinite mass, for simplicity. However, generalization to two-dimensional motions of the vanes and the pawl does not change our main conclusion.
\bibitem{2d2} Allowing the horizontal motion of the pawl, the blocking collision also induces some energy transfer, which
eventually turns into additional heat dissipation into the heat bath 2. Nevertheless, this additional heat dissipation
should be also similar to the form of equation (\ref{eq:Q_int}), which can be made also negligible by taking
the $m_p/m \rightarrow 0$ limit.
\bibitem{overcoming} In principle, $\Delta Q_\textrm{hop}$ can be larger than $U_0$, because the pawl can be lifted to
$y>h$ by the collision with the vanes of energy higher than $U_0+Fx_0$. However, for large $U_0/T_1$, this probability
is exponentially small, so it is sufficient to set $\Delta Q_\textrm{hop} =U_0$ with exponentially small corrections.
\bibitem{back} This is one of the important differences between our case and the original FSR discussed in references~\cite{Feynman,Parrondo}.
\bibitem{fsr} We note that $\langle \dot{Q}_\textrm{col} \rangle_s$ does not vanish in the limit of $m_p/m\rightarrow 0$,
if the interaction bewteen the vanes and the pawl is governed  by a harmonic potential (instead of mechanical collisions in our setup)~\cite{CN}. Hence, in this harmonic case, the Carnot efficiency can not be reached, as claimed by
Parrondo and Espa\~{n}ol~\cite{Parrondo}. A more realistic interaction potential was also considered in the
reference~\cite{Magnasco}.
\bibitem{rev} Feynman assumed in his original FSR~\cite{Feynman} that $\langle \dot{Q}_2 \rangle_s = (r_\textrm{f}-r_\textrm{b})U_0$,
which makes the efficiency equation (\ref{eq:eta}) trivial, independent of rates. In this case, the maximum efficiency is
simply obtained at $F x_0/T_1=\eta_C U_0 /T_2$ by the thermodynamic 2nd law, which gives rise to the Carnot efficiency $\eta_C$.
At this value of $Fx_0$, the entropy production $\Delta S=0$, which indicates that the process is reversible. However, this assumption was criticized by Parrondo and Espa\~{n}ol~\cite{Parrondo}.
\end{thebibliography}
\end{document}

{\huge Supplementary Information for Carnot efficiency is reachable in an irreversible process}

\vskip 2.6cm 

\newcommand{\red}[1]{{\color{red}#1}}

\flushbottom
%\maketitle
\thispagestyle{empty}

\section*{1. Transferred energy per one elastic collision}

\begin{figure*}
\centering
\includegraphics[width=0.4\linewidth]{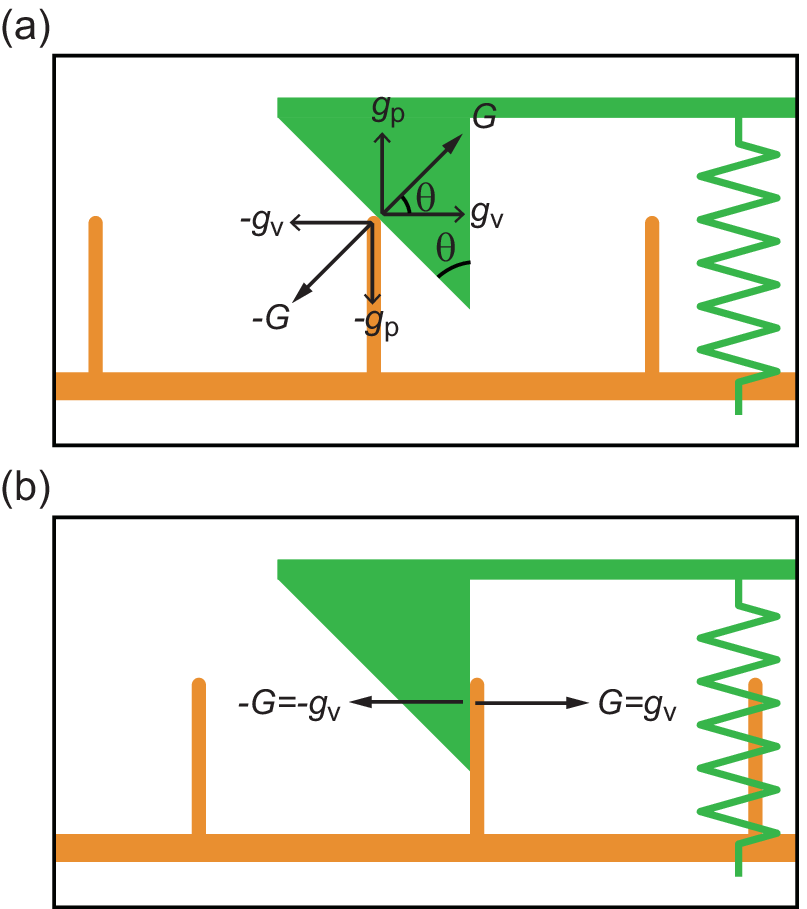}
\caption{ Schematic of the elastic collision model. (a) Collision occurred when vanes move in the forward direction.
When a vane and the pawl collide together, action and reaction forces denoted by $G$ and $-G$ acting on the pawl and a vane respectively, which are orthogonal to the inclined plane of the pawl, are produced. $g_\textrm{v} (= G \cos \theta)$ and $g_\textrm{p} (= G \sin \theta)$ are its horizontal and vertical components, respectively. Since only vertical (horizontal) motion is allowed for the pawl (vanes) by fixed boundary conditions, the horizontal (vertical) component of $G$ does not affect the motion of the pawl (vanes). So, motion of the pawl (vanes) is affected only by $g_\textrm{p}$ ($g_\textrm{v}$) as described in Eq.~(5) (Eq.~(4)) in the main text.
Collision occurs almost instantaneously, so the velocity change during collision are governed by action-reaction forces only:
$m\dot{v} \approx -g_\textrm{v}$ and $m_\textrm{p}\dot{u} \approx g_\textrm{p}$.
 With $\theta = 45^\circ$, we get $m\dot{v} + m_\textrm{p}\dot{u} \approx 0$, which looks like a momentum-conservation law in one dimensional motion of two particles with mass $m$ and $m_\textrm{p}$.
Assuming the elastic collision, we have the energy conservation equation as
$\frac{d}{dt}[mv^2/2 + m_\textrm{p} u^2/2]=0$.
Using these two equations, we can calculate velocities after collision without any detailed knowledge of action-reaction forces.
 (b) Collision occurred when vanes move in the backward direction. In this case, only horizontal forces are produced, which does not affect the motion of the pawl, i.e., $g_\textrm{p} =0$.
To the vane, $g_\textrm{v}$ produces an infinite potential barrier.
} \label{exFig:model}
\end{figure*}

Define $v$ and $v^\prime$ ($u$ and $u^\prime$) as velocities of the vanes (pawl) right before and after the collision as illustrated in Fig.~\ref{exFig:model} (a). From the {\em momentum} and energy conservations as explained in the figure legend, we have
\begin{eqnarray}
\textrm{momentum conservation: }&& mv + m_p u = mv^\prime + m_p u^\prime,  \\
\textrm{energy conservation: }&& \frac{1}{2} mv^2 + \frac{1}{2} m_p u^2
= \frac{1}{2} mv^{\prime 2} + \frac{1}{2} m_p u^{\prime 2}   .
\end{eqnarray}
For a given initial velocities $v$ and $u$, the final velocities $v^\prime$ and $u^\prime$ become
\begin{eqnarray}
v^\prime&& = \frac{(m-m_p) v + 2m_p u  }{m+m_p}, \\
u^\prime&& = \frac{-(m-m_p) u + 2m v  }{m+m_p}.
\end{eqnarray}
Then, the energy transferred by a collision from the vane to the pawl is given by
\begin{equation}
\Delta E \equiv \frac{1}{2}m(v^2-v^{\prime 2}) =
\frac{2 m_p / m}{(1+m_p/m)^2} \left[ m v^2 -m_p u^2 + (m_p - m) v u \right].\label{eq:e-trans}
\end{equation}
Averaging over equilibrium probability distribution functions of velocities, we get
\begin{equation}
\langle \Delta E  \rangle \approx \frac{2m_p/m}{(1+m_p/m)^2}
k_B \left(  T_1 - T_2 \right),  \label{eq:heat_trans}
\end{equation}
where we used $\langle mv^2 \rangle \approx k_B T_1 $ and $\langle m_p u^2 \rangle \approx k_B T_2 $, because
the pawl and the vanes are near in equilibrium with thermal reservoirs  $T_1$ and $T_2$, respectively, in the
high energy barrier limit. The average value of the last correlation term $\langle v u \rangle$ in equation~\eqref{eq:e-trans}  should be exponentially small compared to the first two terms.

%Equation~\eqref{eq:heat_trans} is the origin of heat flow due to collision, defined as $\langle Q_\textrm{col}\rangle_s$ in the main text. Here, the
%heat conductivity is proportional to $m_p/m$ in this elastic model, so we can always make $\langle Q_\textrm{col} \rangle_s \ll \langle Q_\textrm{hop} \rangle_s $ by taking an appropriately small value of $m_p/m$. Note that the numerically obtained exponent value $0.27$ (not $1$) in Fig. 3(e) may come from collision frequency which cannot be easily calculated in our setup.

\begin{figure*}[b]
\centering
\includegraphics[width=0.5\linewidth]{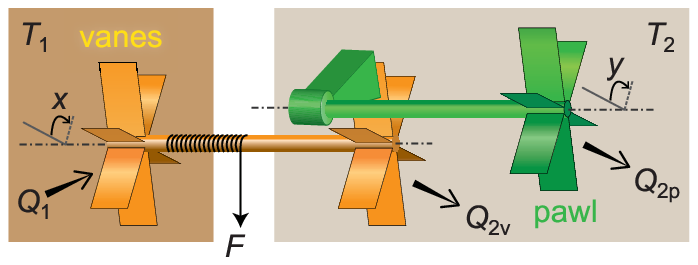}
\caption{Schematic of the original FSR setup.
The vanes are in contact with two baths simultaneously in this setup. Thus, even without the pawl,
irreversible heat $Q_\textrm{2v}$ flows via vanes from the hotter to the colder reservoirs. This additional heat
flow is the crucial element in the argument of Parrondo and Espa\~{n}ol~\cite{Parrondo}, which makes the Carnot efficiency
impossible in this original FSR setup. In contrast, the vanes are in contact with a single heat bath only in our setup
of Fig.~1 (a), so $Q_\textrm{2v}$ simply does not exist. In the presence of the pawl, the mechanical collisions
between the vanes and the pawl will transfer the energy from the hot reservoir into the cold reservoir, denoted
by $Q_\textrm{2p}$, which is composed of two terms as $Q_\textrm{2p}=Q_\textrm{col}+Q_\textrm{hop}$.
}
%call the FSRs of the main text (Fig. 1a) and this Supplementary Information (Fig. S1) as type-I and type-II FSRs, respectively.
%In type-II FSR vanes are affected by two heat baths simultaneously, while vanes are in contact with a single heat bath in type-I FSR. Due to this simultaneous contact, heat  can flow via vanes regardless of the mechanical interaction with the pawl. So, $Q_2 = Q_\textrm{2v}+Q_\textrm{2p}$, where $Q_\textrm{2p}(=Q_\textrm{col}+Q_\textrm{hop})$ is heat transferred from the interaction between the vanes and the pawl.}
\label{exFig:type-II}
\end{figure*}

%\begin{figure*}
%\centering
%\includegraphics[width=0.5\linewidth]{exFig3.eps}
%\caption{ Plot of the efficiency as a function of $F$. The horizontal axis is rescaled as $Fx_0 U_0^{-1} (T_1/T_2-1)^{-1}
%= 1-\alpha/\alpha_m$. Here, $T_1=2$ and $T_2 =1$, thus, $\eta_\textrm{C}=0.5$. As $U_0=5$ is rather small, many forward hops occur as well as a lot of collision events.
%At finite $m_\textrm{p}/m$, $\langle \dot{Q}_\textrm{col} \rangle_s$ is not negligible and affects the efficiency negatively.
%The solid line is drawn by Eq.~(14) in the main text at the value of $U_0=5$, which should be correct only in the %$U_0\rightarrow\infty$ and
%$m_\textrm{p}/m\rightarrow 0$ limits. Thus, this line is not the asymptotic line as
%$m_\textrm{p}/m$ approaches zero, but can be used as an upper-bound guide line.
%As expected, the efficiency increases and seems to approach the vicinity of the guide line as $m_\textrm{p}/m$ decreases. This strongly supports that  our theory works well. One data point is obtained from averaging efficiencies in $10$ steady states and we used $\tau = 2 \times 10^{11}$.
%} \label{exFig:efficiency}
%\end{figure*}

% Bibliography

\vfil\eject